\documentclass[10pt,conference]{IEEEtran}
\IEEEoverridecommandlockouts
\usepackage{cite}
\usepackage{amsmath,amssymb,amsfonts,bm}
\usepackage{algorithmic}
\usepackage{graphicx}
\usepackage{textcomp}
\usepackage{xcolor}
\usepackage{cuted}
\usepackage[font=small,tableposition=top]{caption}
\usepackage[braket]{qcircuit}
\usepackage{lscape}
\usepackage{pgfplotstable,filecontents}
\usepackage[T1]{fontenc}
\usepackage{booktabs}
\usepackage{longtable}
\usepackage{tabu}
\usepackage{url}
\usepackage{breakurl}
\usepackage[breaklinks]{hyperref}
\usepackage{midfloat}

\hypersetup{
  breaklinks=true,
  colorlinks=true,
  linkcolor=blue,
  urlcolor=blue,
  citecolor=blue
}
\pgfplotsset{compat=1.17}
\pgfplotstableset{
    begin table=\begin{longtable},
    end table=\end{longtable},
}
\def\BibTeX{{\rm B\kern-.05em{\sc i\kern-.025em b}\kern-.08em
    T\kern-.1667em\lower.7ex\hbox{E}\kern-.125emX}}

\makeatletter
\newcommand{\linebreakand}{%
  \end{@IEEEauthorhalign}
  \hfill\mbox{}\par
  \mbox{}\hfill\begin{@IEEEauthorhalign}
}
\makeatother

\newcommand{\beq}[0]{\begin{equation}}
\newcommand{\eeq}[0]{\end{equation}}

\begin{document}

\title{QPLEX: Realizing the Integration of Quantum Computing into Combinatorial Optimization Software}

\author{

\IEEEauthorblockN{Juan Giraldo\textsuperscript{*}}
\IEEEauthorblockA{\textit{Department of Computer Science} \\
\textit{University of Victoria}\\
 Victoria, Canada\\
juangiraldo@uvic.ca}

\and

\IEEEauthorblockN{José Ossorio\textsuperscript{*}}
\IEEEauthorblockA{\textit{Department of Computer Science} \\
\textit{University of Victoria}\\
Victoria, Canada\\
joseo@uvic.ca}

\and

\IEEEauthorblockN{Norha M. Villegas}
\IEEEauthorblockA{\textit{Department of Computer Science} \\
\textit{Universidad Icesi}\\
Cali, Colombia\\
nvillega@icesi.edu.co}

\and

\linebreakand

\IEEEauthorblockN{Gabriel Tamura}
\IEEEauthorblockA{\textit{Department of Computer Science} \\
\textit{Universidad Icesi}\\
Cali, Colombia\\
gtamura@icesi.edu.co}

\and

\IEEEauthorblockN{Ulrike Stege}
\IEEEauthorblockA{\textit{Department of Computer Science} \\
\textit{University of Victoria}\\
Victoria, Canada\\
ustege@uvic.ca}
}

\maketitle

\begingroup\renewcommand\thefootnote{*}
\footnotetext{These authors contributed equally to this work.}
\endgroup

\begin{abstract}
Quantum computing has the potential to surpass the capabilities of current classical computers when solving complex problems. Combinatorial optimization has emerged as one of the key target areas for quantum computers as problems found in this field play a critical role in many different industrial application sectors (e.g., enhancing manufacturing operations or improving decision processes). Currently, there are different types of high-performance optimization software (e.g., ILOG CPLEX and Gurobi) that support engineers and scientists in solving optimization problems using classical computers. In order to utilize quantum resources, users require domain-specific knowledge of quantum algorithms, SDKs and libraries, which can be a limiting factor for any practitioner who wants to integrate this technology into their workflows. Our goal is to add software infrastructure to a classical optimization package so that application developers can interface with quantum platforms readily when setting up their workflows. This paper presents a tool for the seamless utilization of quantum resources through a classical interface. Our approach consists of a Python library extension that provides a backend to facilitate access to multiple quantum providers. Our pipeline enables optimization software developers to experiment with quantum resources selectively and assess performance improvements of hybrid quantum-classical optimization solutions.
\end{abstract}


\section{Introduction}

Optimization plays a key role in industry, solving computationally challenging problems that arise in multiple fields, such as logistics \cite{geunes_pardalos_2005}, manufacturing \cite{gupta_gupta_2020}, finance \cite{cornuejols_peña_tütüncü_2018}, among others. Given the complexity of these problems, optimization software is often needed to facilitate their modeling and analysis. This type of software is commonly provided through programming libraries such as DOcplex \cite{docplexibm} (a Python wrapper for ILOG CPLEX \cite{manual1987ibm}), which deliver a set of functionalities for the solution of optimization problems through the use of classical backends, also known as high-performance solvers. These backends are capable of solving linear, mixed integer and quadratic programming problems, making them very versatile for different optimization applications. Nonetheless, such classical approaches come with certain limitations regarding the size of the problem and the complexity of the objective function and their constraints \cite{venter2010}.

In the past few years, quantum computers have become widely available \cite{9605329}, allowing the scientific community to come up with near-time quantum solutions for complex problems by exploiting the principles of quantum mechanics to achieve a computational advantage when compared to their classical counterparts \cite{Golden}, \cite{farhi}, \cite{Ambainis1807.05209}. Optimization sits at the core of many of these problems, making it one of the areas that might benefit from quantum computing. Recent advances in this field have provided techniques for solving complex optimization problems through different paradigms such as adiabatic \cite{Mukherjee2015}, \cite{10.1145/2482767.2482797}, \cite{Rieffel_2014} and gate-based quantum computing \cite{Liu_2022}, \cite{8728102}, \cite{9810536}, \cite{1708.05294}. These techniques could potentially bring a speed up compared to classical approaches, while delivering higher quality solutions.

Given the variety of quantum computing providers and algorithm implementations for solving optimization problems, some techniques executed on certain devices may yield better results, achieve better performance, or be more efficient than others for a given problem \cite{9605329}. Thus, testing different algorithmic approaches and quantum processing units is essential in order to find the best alternative for a specific use case. This is especially important if one intends to benchmark against classical optimizers.

Currently, each quantum hardware provider uses its own library or SDK to allow users to interact with their machines: Qiskit \cite{Abraham2019-xj} in the case of IBM, Ocean \cite{D-Wave_Systems_Inc_undated-pf} for D-wave and Pennylane \cite{pennylane} for Xanadu, to name a few. This situation presents a challenge for any developers who want to implement an optimization solution for multiple quantum devices as they would have to adapt the code to function with each target platform. Moreover, solving optimization problems using quantum computers is not a trivial task; a certain level of proficiency in quantum algorithms and programming is necessary to develop such solutions. This adds friction to the programming experience, making it considerably more difficult for a software engineer to start experimenting with quantum computing.

Addressing these pain points, we conceived and developed QPLEX, a Python library extension based on DOcplex that allows developers to implement a general mathematical formulation once (also known as an optimization model) and execute it seamlessly on multiple quantum devices using different quantum algorithms. Our solution automatically handles the adaptation of a general optimization model into the specific instructions used by the target quantum device’s SDK. Additionally, the library offers a versatile gate-based algorithm execution workflow that is capable of running two gate-based quantum algorithms for optimization (i.e. QAOA \cite{farhi2014quantum} and VQE \cite{Peruzzo_2014}) in a hardware-agnostic manner. This approach enables seamless execution on any supported device without incurring extra programming overhead.

This paper is divided into two main sections. First we review current techniques for solving optimization problems with quantum computers and discuss background and related work on quantum optimization. Second, we describe our approach and methods and present future directions for this project. Lastly, we go over our contributions and layout improvements for our project as future work.

\section{Quantum Optimization Algorithms}
The three algorithmic methods currently most used in quantum optimization are quantum annealing (QA), quantum approximate optimization algorithm (QAOA), and variational quantum eigensolver (VQE), which we describe in subsections \ref{sec/quantum-annealing}, \ref{sec/QAOA}, and \ref{sec/VQE}, respectively. These algorithms have been proven to be fitting options for solving optimization problems using near-term quantum computers \cite{Preskill2018} in combination with classical processing. A recent study \cite{mugel2022dynamic} implemented QA and VQE to solve the dynamic portfolio optimization problem with real data, obtaining promising results with hybrid quantum-classical approaches.

Our solution, QPLEX, supports solving an optimization model using QA, QAOA and VQE. The implementation of the QA algorithm, as we describe, requires small adjustments to the QPLEX model to be executed on a quantum annealer. However, implementations of QAOA and VQE need some adaptation to translate the general optimization model into a quantum circuit. This functionality is made possible by the Generalized Gate-Based Algorithm Execution (GGAE) workflow, which we describe in section \ref{sec/GAAR}.

To the best of our knowledge, currently there is no such tool that allows for the seamless execution of a general representation of an optimization problem using various different algorithms and quantum providers.

\subsection{Quantum Annealing} \label{sec/quantum-annealing}
Quantum annealing is a quantum algorithm that uses adiabatic quantum computation. This algorithm is fitting for solving optimization problems due to the fact that the general model can be represented in a quadratic unconstrained binary optimization (QUBO) form, which can be translated into a Hamiltonian or energy function, turning the problem into an energy minimization problem that can be solved with ease by a quantum annealer. D-Wave provides a hybrid solver that takes a problem and breaks it into smaller parts that are solved using quantum-classical strategies. This approach allows the machine to handle problem instances that would normally be beyond its capabilities if it only used quantum resources \cite{mugel2022dynamic} \cite{DWaveAnnealing}.

\subsection{Variational Quantum Eigensolver} \label{sec/VQE}
Another approach for solving optimization problems is the Variational Quantum Eigensolver \cite{Peruzzo_2014}, which is a hybrid quantum-classical algorithm that approximates the ground state of the Hamiltonian representing the optimization problem \cite{mugel2022dynamic}. The eigenvalues of the Hamiltonian are the possible measurements of the quantum system at each of the eigenstates. The algorithm works by sampling the function distribution to estimate the energy of the quantum state. The starting state, known as the ansatz, is implemented by parameterized circuits, and after each estimation the circuit's variational parameters are optimized by classical techniques. Eventually, the energy estimation converges to the ground state of the Hamiltonian, the lowest eigenstate, with its eigenvalue as the solution to our optimization problem. The quality of the results of this algorithm depend first and foremost on the variational quantum circuit used.

\subsection{Quantum Approximate Optimization Algorithm} \label{sec/QAOA}

The Quantum Approximate Optimization Algorithm (QAOA) is a hybrid quantum-classical algorithm specifically designed for solving combinatorial optimization problems \cite{farhi2014quantum}. Similarly to VQE, this algorithm also uses a parameterized circuit as an anzatz to encode the problem. Nevertheless, in the case of QAOA, this initial state is problem-specifc and constructed using an alternating combination of a problem Hamiltonian in the form of a QUBO and a mixer Hamiltonian. Both of these sections are repeated a $p$ number of times to improve the level of approximation of the algorithm. The use of a large value of $p$ comes with a trade off as this increases the complexity circuit depth and therefore the execution time. A classical optimization routine is employed to find the best combination of parameters for the initial circuit state. The execution time of this step is also determined by $p$, as the more layers the algorithm has the more parameters need to be optimized. In addition to the original QAOA implementation, different extensions have been developed for achieving better approximations. For instance, the Quantum Alternating Operator Ansatz \cite{Hadfield_2019} achieves this by allowing the use of more general alternating mixer operators. The Warm-start QAOA \cite{Egger_2021} includes an additional step to solve the problem approximately with a different optimization method (e.g., a classical solver) and then use that output as the initial state to QAOA.

\section{QPLEX Programming Library}
QPLEX facilitates the design and execution of optimization models on quantum hardware, enabling users to access quantum resources in an effective way. This, in turn, allows to perform comparisons between classical and quantum solvers conveniently. To accomplish this goal, the QPLEX programming library is equipped with a set of important features that allow for (1) the representation of an optimization problem, (2) the adaptation of the optimization model to different supported quantum processing units (QPUs), and (3) the building of a hardware-agnostic algorithmic implementation based on the model requirements. In order to understand how the library works, we expand on each of these features in the subsequent subsections.
\subsection{Building a general representation} \label{sec/qplex-model}
\begin{figure}[h]
\centering
\includegraphics[width=0.5\textwidth]{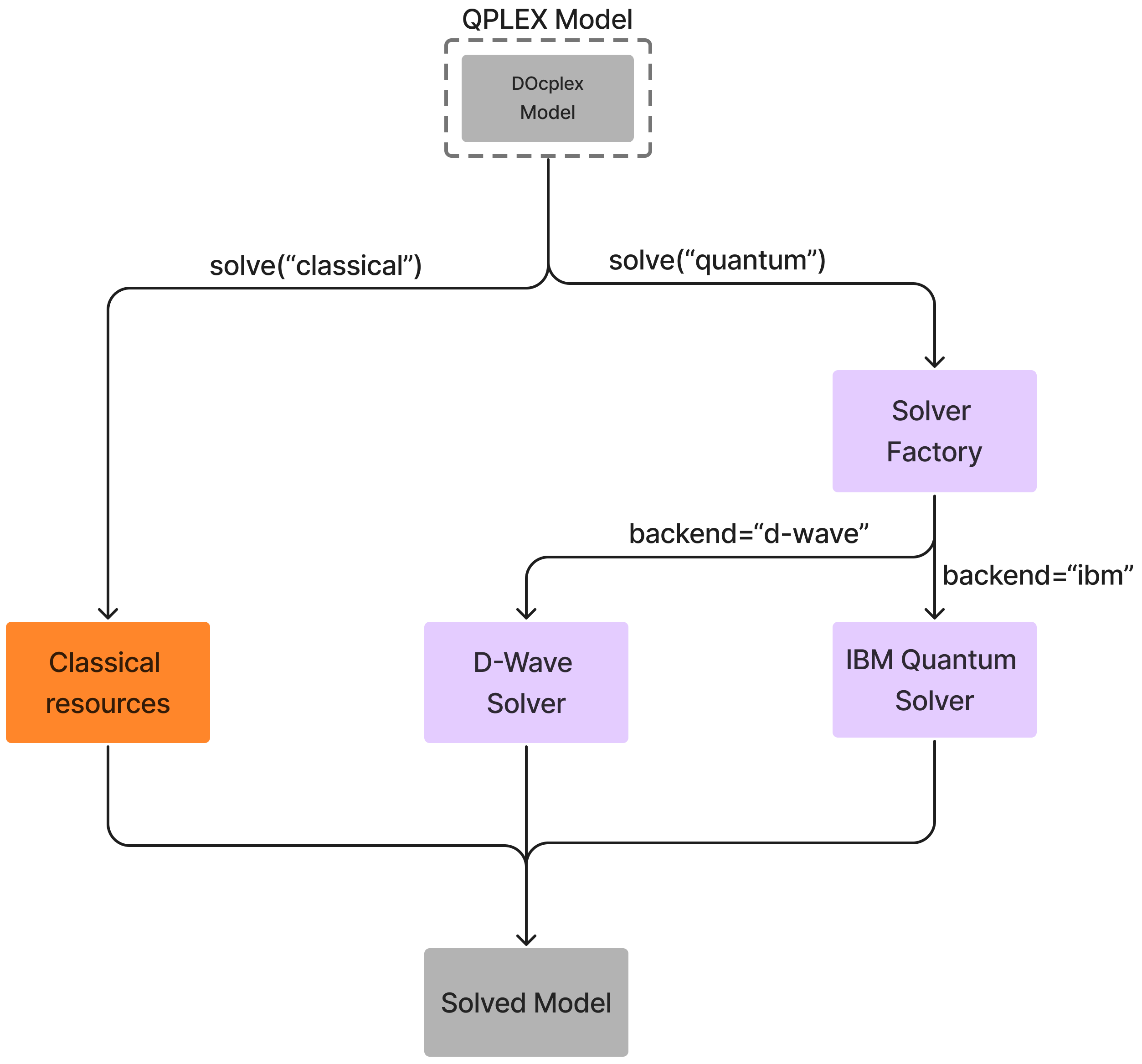}
\caption{QPLEX library workflow} \label{fig/qplex-workflow}
\end{figure}
To accomplish a general representation of an optimization problem that can be executed on different quantum hardware, we devised a wrapper for a base DOcplex model which encapsulates the default behavior of the classical high-performance solver (CPLEX) and the customized workflow execution for the supported quantum providers (IBM and D-Wave). The newly conceived model contains the same attributes and methods as a DOcplex model. Nevertheless, the QPLEX model class overrides the \textit{solve} method, allowing the user to specify the desired solver to be utilized (i.e., classical or quantum) and, in the quantum case, to select a quantum provider. 

Given that DOcplex is used as an API to model the combinatorial optimization formulations, it was necessary to allow for the execution of any type of optimization problem within QPLEX. These include unconstrained and constrained (i.e., equality and inequality constraints) problems, as well as the use of binary, discrete, and continuous variables. Taking into account that the supported quantum algorithms for combinatorial optimization require the problem to be in the form of a QUBO formulation, it was necessary to devise a method for converting non-compliant formulations when needed. For practicality purposes, Qiskit's QUBO converters were used to address this situation. However, we plan to implement our own mapping and conversion methods in the near future.

Figure \ref{fig/qplex-workflow} illustrates the QPLEX model workflow. On the one hand, if the classical solver is selected,  the program will simply call the base model's \textit{solve} method, which in turn executes the underlying problem formulation using the CPLEX optimizer. On the other hand, if the quantum solver is selected, the model will use the specified backend to execute the problem on quantum hardware through their corresponding cloud (e.g., D-Wave Leap, or IBM Quantum). Note that for the user the only difference is the choice of the parameters of the \textit{solve} method since the underlying solving process is handled automatically by the library, and the solved model is accessible in the same manner as the base model. In other words, the switch between using quantum or classical resources to solve the problem is practically seamless for the developer.

The following code snippet illustrates the use of QPLEX to execute an optimization problem model (the knapsack problem) on quantum resources while using the same syntax provided by DOcplex:

\begin{figure}[h]
\centering
\includegraphics[width=0.5\textwidth]{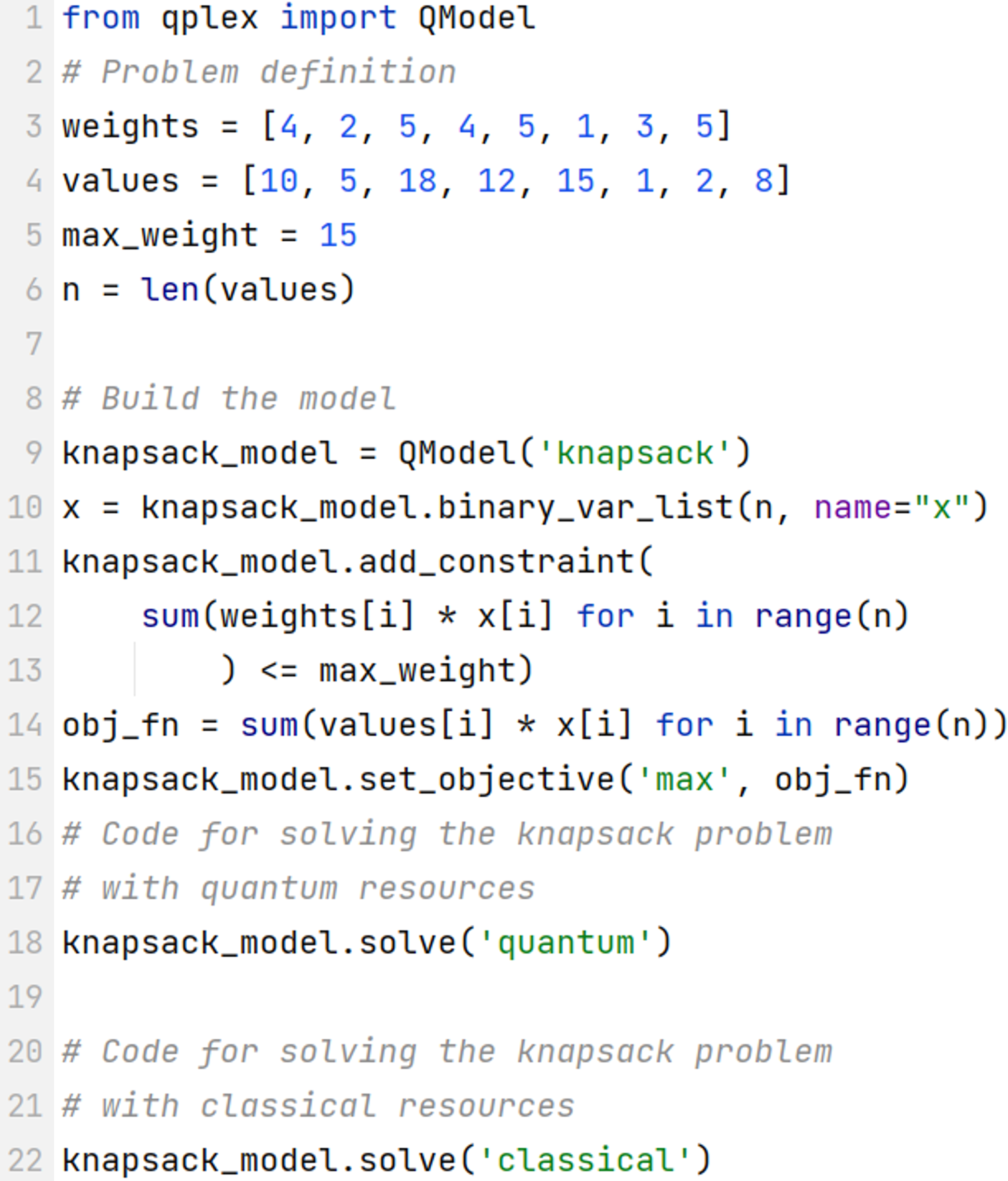}
\caption{Code snippet for solving the knapsack problem with QPLEX} \label{fig/qplex-snippet}
\end{figure}

Moreover, some configuration parameters of the solve method are only necessary in specific cases. For instance, specifying a quantum algorithm to solve the problem would only make sense in the case that the user has selected a gate-based quantum provider.

\subsection{Optimization model adaptation for different QPUs} \label{sec/quantum-adapter}

One of the primary characteristics of QPLEX is its capability to translate a general representation of an optimization problem into the particular set of instructions employed by the quantum providers available in the programming library for addressing the desired problem formulation on their QPUs. This process is realized through multiple solver instances. Each solver corresponds to a quantum backend (e.g. IBM, D-Wave) and they abstract vital functionalities for the adaptation. For instance, how to parse the problem formulation, how to execute it, and how to read the results sent back from the QPU. The problem execution includes the quantum hardware selection stage where, if multiple devices are available for use, only the ones with a sufficient number of qubits to support the formulation and the shortest job queue are considered. The solver handles the access to quantum computing resources through requests to its corresponding backend's cloud. These services require an API token to successfully execute the problem, making it necessary for QPLEX users to have their own tokens and set them as environment variables for the library to function correctly.

\begin{figure}[h]
\centering
\includegraphics[width=0.5\textwidth]{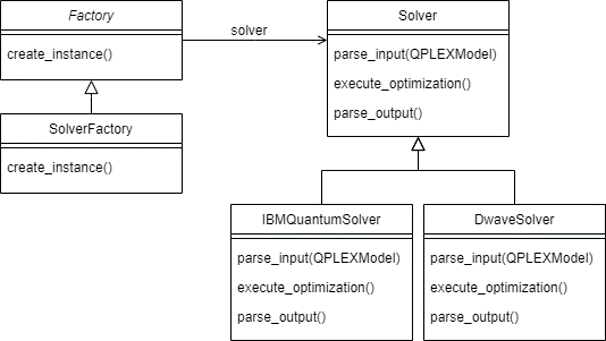}
\caption{Quantum Solver Factory} \label{fig/adapter}
\end{figure}

\begin{figure}[h]
\centering
\includegraphics[width=0.5\textwidth]{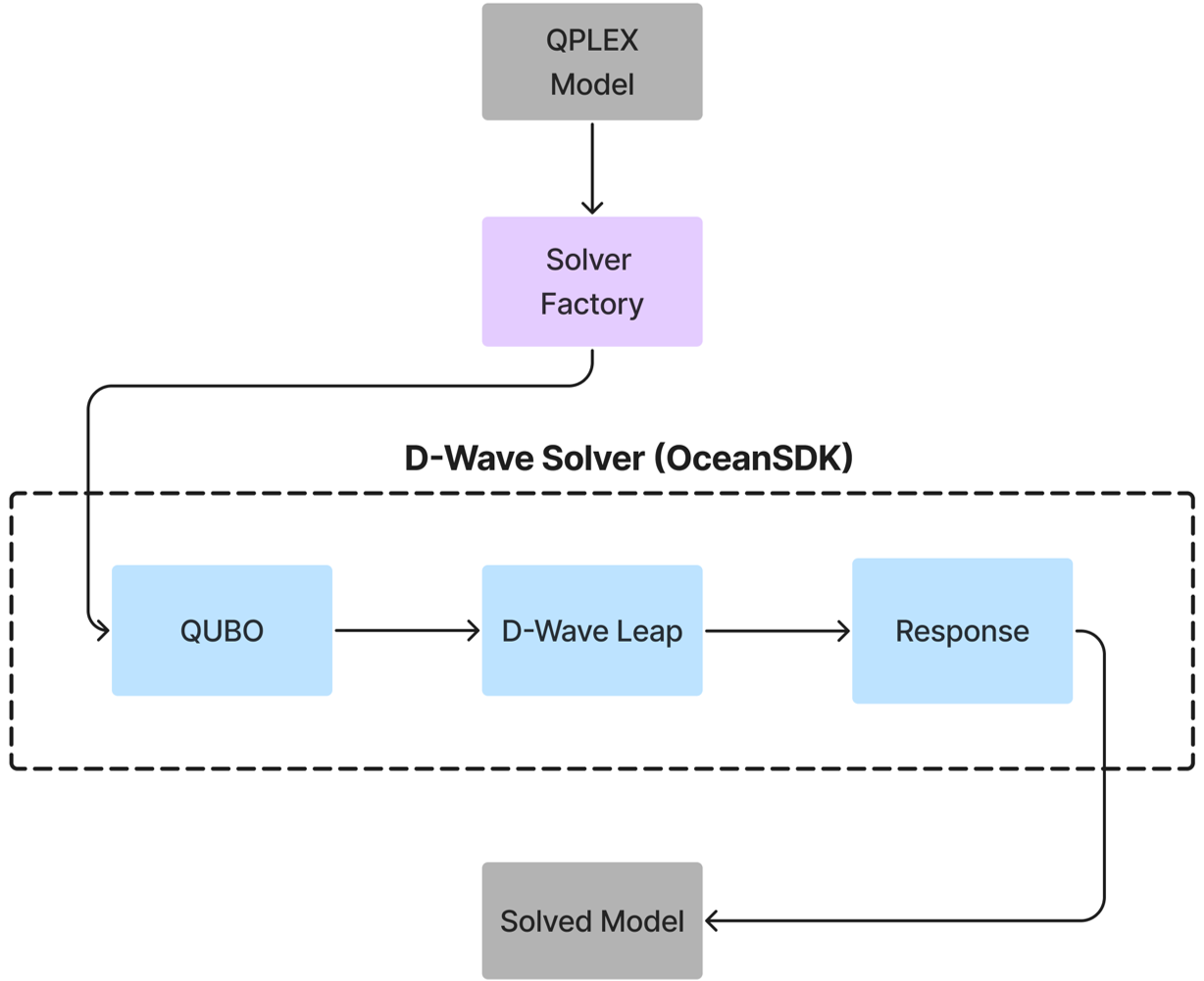}
\caption{Model execution using the D-Wave solver} \label{fig/d-wave-solver}
\end{figure}

In order to manage the creation of quantum solver instances, a solver factory module is implemented. Depending on the selected hardware provider, this module returns the corresponding solver for the optimization problem to be addressed. This design is  depicted in Figure \ref{fig/adapter}. All QPLEX models instantiate the solver factory and make use of it when a quantum backend is selected, allowing the model to understand which set of instructions has to be executed to solve the specified optimization problem. Figure \ref{fig/d-wave-solver} illustrates the methods described above using the D-Wave solver as a specific example. In this case, the factory builds a solver instance that leverages instructions written in the OceanSDK to convert the base model into a QUBO and execute it on a D-Wave Leap device. The machine's response is then parsed and returned as the solution for the optimization problem.

In addition to providing a simple workflow for execution, the design of this module greatly minimizes the coupling between the solvers and the rest of the system, as all the logic for each of these is contained within its own class and used through a common call from the factory. In the case that a quantum provider decides to update how to access their machines, it is only necessary to modify the corresponding solver class; the remaining sections of code are not affected by the change. This also applies when a new backend has to be integrated into the library, as it is only needed to create the new solver class with the necessary execution logic and instantiate it within the solver factory, heavily reducing the amount of time and complexity required to provide access to a new quantum provider.

\subsection{Generalized Gate-Based Algorithm Execution} \label{sec/GAAR}

As mentioned before, quantum annealers are designed to solve problem formulations in QUBO forms by directly mapping them into a QPU. This allows for an effective execution workflow for optimization problems as the same QUBO can be solved in different annealers using their specific SDK or programming library. Nevertheless, this is not the case for general-purpose gate-based quantum computers. In order to solve an optimization problem through one of these machines, it is necessary to first implement a quantum algorithm to approach the optimization task. Most of the time these implementations are provider specific, which makes it necessary to have multiple versions of the same approach when experimenting on different quantum computers. Thus, adding support for more devices and quantum algorithms is cumbersome.

\begin{figure}[h]
\centering
\includegraphics[width=0.5\textwidth]{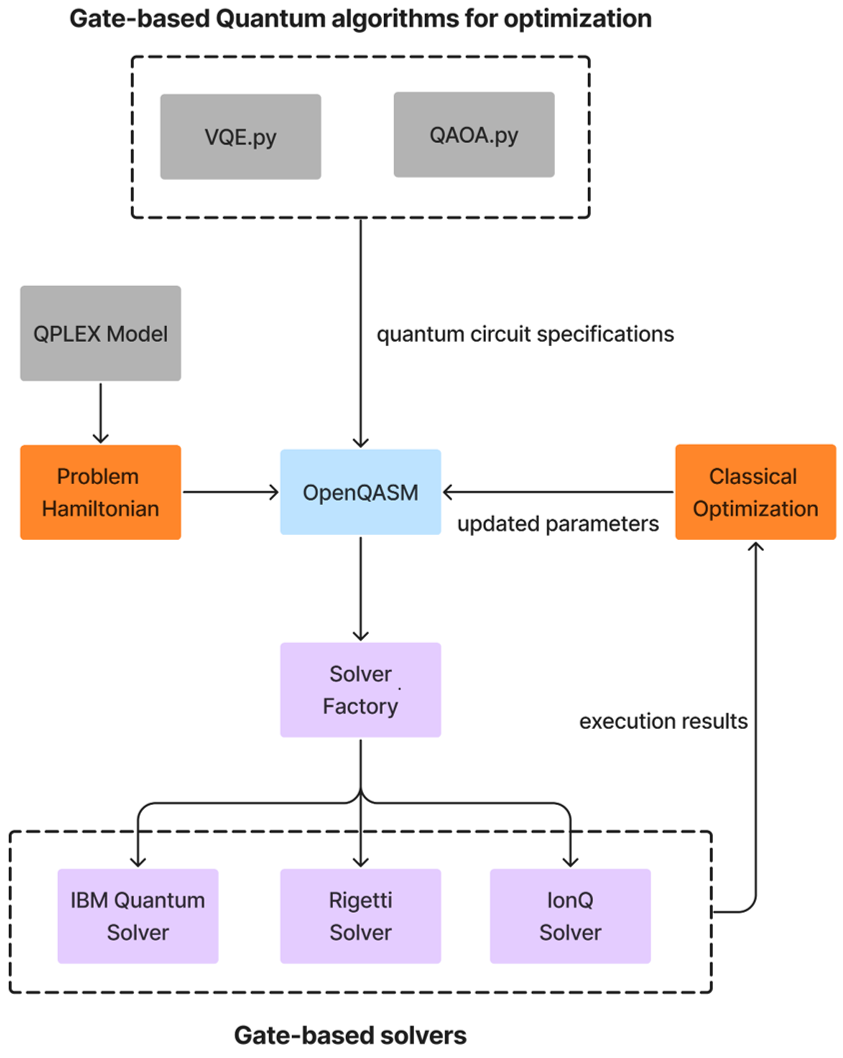}
\caption{The generalized gate-based algorithm execution workflow} \label{fig/GGAR}
\end{figure}

To improve this situation, we developed a workflow for the execution of hardware-agnostic gate-based quantum algorithms for optimization, depicted in Figure \ref{fig/GGAR}. This process begins once the optimization model has been created and the user has chosen to solve the problem through a gate-based quantum machine. The first step in this approach is to transform the QPLEX model into a QUBO formulation that can be solved employing any of the variational algorithms provided by our library.

As it was previously described, this type of algorithm comprises two main phases: the quantum circuit execution and the classical parameter optimization loop. The generated QUBO model is employed in the first phase to build the initial parameterized quantum circuit or anzatz. Given that our main goal is to make this execution hardware-agnostic, the circuit is constructed using OpenQASM3 directives instead of SDK specific quantum gates. Since the execution is going to be determined by this intermediate representation language, it is required that the gate-based solvers support OpenQASM3, as once the generalized circuit is created, each backend has to transpile the instructions into QPU specific operations.

Once the operations are executed by the appropriate solver, the solution for the first iteration of the algorithm is then returned to be used in the optimization loop. For this step, a classical optimization algorithm (e.g., SPSA, COBYLA or Adam) is used in conjunction with a cost function, to find the optimal configuration of parameters for the circuit. It is possible for the user to specify which optimizer to be employed, but if none is provided COBYLA will be used by default. With the optimization results a new generalized circuit is constructed and once again executed through a solver. The loop is performed a set number of times or until the loss stops changing. At this point the best solution is returned to the user in a seamless manner.

The discussed setup allows for a simple way to incorporate new implementations of quantum optimization algorithms, as the general instructions for the implementation have to be specified only once and can be re-used for multiple gate-based solvers.

\section{Conclusions and Future Work}
In this paper, we have presented QPLEX, a quantum hardware-agnostic Python library for optimization, detailing its motivation, design, implementation, and usability. We introduced the primary components of our approach, specifically the QPLEX model and the quantum solver factory, and demonstrated their interaction in facilitating a seamless workflow for solving optimization problems effectively in quantum infrastructure. Our solution builds on top of a known optimization library (i.e., DOcplex) and abstracts the syntax related to the different quantum SDKs needed to communicate with different quantum backends, as well as the necessary parsing of a general optimization model defined using DOcplex to the specific quantum target's language. We also show the flexibility of this toolbox to be extended and adapted to specific development needs, making it a useful tool for a wide range of scenarios. 

Moreover, we would like to highlight our approach of using a common programmatic interface as a means to provide access to a large collection of quantum algorithms and hardware providers as one of the main contribution of this work. While packages for quantum optimization currently exist, this paper presents, to the best of our knowledge, the first attempt to unify the execution of classical and quantum, both annealing and gate-based techniques for optimization problems under the same development tool.

For this paper, we only incorporated three quantum algorithms (i.e., Quantum Annealing, VQE, QAOA) and two providers (i.e., IBM and D-Wave) into QPLEX; nonetheless, moving forward, we expect to include more algorithms and support access to a larger number of quantum computers. Currently, QPLEX is an open-source project available on GitHub \footnote{https://github.com/JuanGiraldo0212/QPLEX}, and we encourage readers who are interested to contribute to any of the open issues, as we intend to make this library a useful tool built and used by the QC community. Finally, despite current limitations in quantum hardware, we believe that the continued development of quantum software engineering solutions is crucial to fully realize the potential benefits of fault-tolerant quantum computing in the near future.

\bibliographystyle{IEEEtran}
\bibliography{IEEEabrv, qse}

\end{document}